\documentclass{article}

\usepackage[english]{babel}
\usepackage[utf8x]{inputenc}
\usepackage[T1]{fontenc}
\usepackage[title]{appendix}
\usepackage{framed}
\usepackage{color}
\usepackage{algorithm}
\usepackage{algpseudocode}
\usepackage{float}
\usepackage{enumerate}
\usepackage{enumitem}
\usepackage[a4paper,top=3cm,bottom=2cm,left=3cm,right=3cm,marginparwidth=1.75cm]{geometry}
\usepackage{amsmath}
\usepackage{amssymb}
\usepackage{amsthm}
\usepackage{proof}
\usepackage{graphicx}
\usepackage{subcaption}
\usepackage[colorinlistoftodos]{todonotes}
\usepackage[colorlinks=true, allcolors=blue]{hyperref}
\usepackage{wrapfig,lipsum}
\usepackage{authblk}
\usepackage{multirow}
\usepackage{hhline}
\usepackage{setspace}
\usepackage{palatino}
\usepackage{mathdots}
\usepackage{multirow}
\usepackage{siunitx}
\usepackage{xcolor}
\usepackage{booktabs,colortbl, array}

\newcolumntype{C}[1]{>{\centering\arraybackslash}p{#1}}

\newcommand\ket[1]{\ensuremath{|#1\rangle}}
\newcommand\bra[1]{\ensuremath{\langle#1|}}

\begin{document}

\title{Alibaba Cloud Quantum Development Platform: Large-Scale Classical Simulation of Quantum Circuits}

\author[1,2]{Fang Zhang}
\author[1,2]{Cupjin Huang}
\author[3]{Michael Newman}
\author[4]{Junjie Cai}
\author[5]{Huanjun Yu}
\author[5]{Zhengxiong Tian}
\author[6]{Bo Yuan}
\author[5]{Haihong Xu}
\author[4]{Junyin Wu}
\author[1]{Xun Gao}
\author[1]{Jianxin Chen\footnote{liangjian.cjx@alibaba-inc.com}} 
\author[1]{Mario Szegedy}
\author[1]{Yaoyun Shi}

\affil[1]{Alibaba Quantum Laboratory, Alibaba Group USA, Bellevue, WA 98004, USA}
\affil[2]{Department of Electrical Engineering and Computer Science, University of Michigan, Ann Arbor, MI 48109, USA}
\affil[3]{Departments of Physics and Electrical and Computer Engineering, Duke University, Durham, NC 27708, USA}
\affil[4]{Alibaba Cloud Intelligence, Alibaba Group USA, Bellevue, WA 98004, USA}
\affil[5]{Alibaba Cloud Intelligence, Alibaba Group, Hangzhou, Zhejiang 310000, China}
\affil[6]{Alibaba Infrastructure Service, Alibaba Group, Hangzhou, Zhejiang 310000, China}
\setcounter{Maxaffil}{0}
\renewcommand\Affilfont{\itshape\small}

\date{\today}
\maketitle

\begin{abstract}
We report, in a sequence of notes, our work on the Alibaba Cloud Quantum Development Platform (AC-QDP). AC-QDP provides a set of tools for aiding the development of both quantum computing algorithms and quantum processors, and is powered by a large-scale classical simulator deployed on Alibaba Cloud.  
In this note, we report the computational experiments demonstrating the classical simulation capability of AC-QDP. We use as a benchmark the random quantum circuits designed for Google's Bristlecone QPU {\cite{GRCS}}. We simulate Bristlecone-70 circuits with depth $1 + 32 + 1$ in $0.43$ second per amplitude, using $1449$ Alibaba Cloud Elastic Computing Service (ECS) instances, each with $88$ Intel Xeon(Skylake) Platinum 8163 vCPU cores @ 2.5 GHz and $160$ gigabytes of memory. By comparison, the previously best reported results for the same tasks are $104$ and $135$ seconds, using NASA's HPC Pleiades and Electra systems, respectively ({arXiv:1811.09599}). Furthermore, we report simulations of Bristlecone-70 with depth $1+36+1$ and depth $1+40+1$ in $5.6$ and $580.7$ seconds per amplitude, respectively. To the best of our knowledge, these are the first successful simulations of instances at these depths. 
\end{abstract}

\medskip

Building quantum computers and developing their applications are the two primary challenges for the field of quantum computing. We are at an early stage for quantum computing that is often likened to the development of classical computers in the early 20th century.
However, there is a fundamental difference: for the design of quantum computers and applications, we now have at our disposal powerful classical computing capabilities that have been exponentially improving for decades. The Alibaba Cloud Quantum Development Platform (AC-QDP) aims to utilize Alibaba's massive classical computational resources for aiding the development of quantum applications and quantum computers themselves. 

The computational engine of AC-QDP is at present our classical quantum circuit simulator Tai-Zhang, deployed on Alibaba Cloud. In~\cite{CZH+18}, we described Tai-Zhang's algorithm and the computational experiments that deployed it on the computing facilities in Alibaba Group's Data Infrastructure and Search Technology Division. For migration to Alibaba Cloud and to adapt to the updated benchmark, we made several technical changes to Tai-Zhang, documented below. We refer the interested reader to~\cite{CZH+18} for the algorithm.

Our choice of benchmark, first proposed in the Google AI Blog~\cite{BN18}, is motivated by the prospect of comparing our results with those of other groups and with a real quantum device, such as the Bristlecone quantum processor Google is working on~\cite{Bristlecone}. Thus, simulating these circuits will provide a direct comparison between quantum and classical implementations for the same task. This has motivated several other groups to simulate these circuits as well.

We recognize, however, that being able to simulate these circuits does not guarantee the ability to simulate other circuits of a similar scale. In particular, like all simulations based on tensor contraction, our approach requires computational resources that scale exponentially in the treewidth of the quantum circuit, and thus is limited fundamentally in the size it can simulate. Nevertheless, as we will report in subsequent notes, our system can be applied fruitfully despite this fundamental constraint.

\section{Benchmarking and the Experimental Setup} \label{sec:setting}
The circuits we simulate in this work are described in~\cite{BN18}. They are a  modification of the random circuits defined in \cite{BIS+16}, and the circuit files are available for download \cite{GRCS}. The new circuit prescription is as follows.

\begin{enumerate}\label{GRCSv1}
\item Begin with a layer of all Hadamard gates.
\item Place $t$ layers of CZ gates alternating between 8 configurations.
\item Place single-qubit gates on these $t$ layers at positions unoccupied by CZ gates, according to the following rules.
\begin{enumerate}
\item Place a single-qubit gate chosen at random from the set $\{\sqrt{\mathrm{X}}, \sqrt{\mathrm{Y}}\}$ at a qubit if that qubit participates in a CZ gate in the previous layer.
\item Place a $\mathrm{T}$-gate at a particular qubit if that qubit participates in a $\sqrt{\mathrm{X}}$, $\sqrt{\mathrm{Y}}$, or $\mathrm{H}$ gate in the previous layer.
\end{enumerate}
\item End with a final layer of all Hadamard gates.
\end{enumerate}

A series of papers address the simulation of these revised random quantum circuits \cite{VBN+18,VLB+19, CLG+19, GLX+19}. In \cite{VBN+18}, the authors estimated that the revised circuits should be about $1000\times$ harder to simulate, compared with those in \cite{CZH+18}. In \cite{VBN+18}, Bristlecone-70 with depth $1+32+1$ has been benchmarked on NASA's HPC Pleiades and Electra systems with reported runtimes of $2.89\times 10^{-2}$ and $3.57 \times 10^{-2}$ hours respectively, or equivalently $104.04$ and $128.52$ seconds. 
For the above computation, all four available node architectures on the Pleiades system are used:
\begin{enumerate}
\item 2016 Broadwell (bro) nodes: Intel Xeon E5-2680v4, 28 cores, 128GB per node;
\item 2088 Haswell (has) nodes: Intel Xeon E5-2680v3, 24 cores, 128GB per node;
\item 5400 Ivy Bridge (ivy) nodes: Intel Xeon E5-2680v2, 20 cores, 64GB per node;
\item 1936 Sandy Bridge (san) nodes: Intel Xeon E5-2670, 16 cores, 32GB per node;
\end{enumerate}
and two available archictures on the Electra system are used:
\begin{enumerate}
\item 1152 Broadwell (bro) nodes: same as above;
\item 2304 Skylake (sky) nodes: $2 \times 20$-core Intel Xeon Gold 6148, 40 cores, 192GB per node.
\end{enumerate}
In \cite{CLG+19}, the same circuit has been argued to be simulable for supercomputers like Tianhe-2 by analyzing the circuit complexity. Additionally, 72-qubit random quantum circuits for Bristlecone with depth $1+32+1$ have been benchmarked in \cite{CLG+19}, which reported runtimes of $14.1$ minutes, or $846$ seconds to compute a single amplitude on $16384$ Sunway SW26010 260C nodes, with $256$ cores each.  

Before we present our simulation setup, we make a few clarifications. First, Bristlecone-70, the 70-qubit random quantum circuit family for that architecture, is equivalent for simulation purposes to Bristlecone-72, the 72-qubit version, since two qubits of the latter network can be easily contracted with their only neighbor. Therefore, we can compare the above reported runtimes with results using Bristlecone-70 at depth $1+32+1$, even though we use less CPU cores than previously reported work.  
Second, we adopt a slightly different notation. In \cite{GRCS}, the file bris\_n.tar.gz contains circuits using $n$ rows. The files inside are named bris\_n\_maxcycle\_id.txt. Bristlecone-m refers to those circuits acting on $m$ qubits. So the 70-qubit circuits that we benchmark in this paper, Bristlecone-70, correspond to circuit description files in Bris\_11.tar.gz.  In particular, we emphasize that these are the largest size circuits available in \cite{GRCS}. 

We use 1449 Alibaba Cloud Elastic Compute Service (ECS) instances, each with 88 virtual CPU cores and 160 GB of memory. We first use a single node as an agent to split the large tensor network contraction task into many smaller tensor network contraction subtasks; this step is called `preprocessing'. Then, the agent node uses the OSS (Object Storage Service) as a data transmission hub to assign different subtasks to different nodes. When a node is finished with the assigned subtask, it will upload the result to the OSS, and the agent node will repeatedly query the OSS until all the subtask results add up to the desired amplitude. The reported `running-time' is the total elapsed time obtained on the agent node except for the preprocessing step. This is because we only need to do preprocessing once, independent of the number of amplitudes we will calculate.

We refactored the source code of the simulator we presented in \cite{CZH+18} to yield a better abstraction of the simulation task. A major change in the refactoring is to switch the edges and the nodes of the tensor network. In the previous version, we denoted the running indices of a tensor network as nodes, while each tensor was regarded as a hyper-edge connecting several nodes together. The refactored code is formulated in the opposite way, where each node now holds a tensor and each hyper-edge corresponds to a running index. This change provides a common interface for all tensor-valued objects, and allows us to conveniently construct complex tensor networks. Despite this change, the algorithm in the refactored code is functionally similar to that in \cite{CZH+18}, and so we don't observe a significant performance difference after the refactorization. For more details of the algorithm, please refer to \cite{CZH+18}. Detailed benchmarking results will be presented in Sec~\ref{sec:result}.

\section{Benchmarking Results} \label{sec:result}

Several variables will affect the performance of our simulator:
\begin{itemize}
\item the number $N_{a}$ of amplitudes to calculate;
\item the number $N_{c}$ of CPU cores;
\item the number $N_{s}$ of subtasks for calculating a single amplitude.
\end{itemize}

In our algorithm, all subtasks have equal computational complexity. Thus, the naive way to balance the computational load is to assign $\frac{{N_a}N_{s}}{N_s}$ subtasks to each CPU core.  There is no shared data access across these subtasks, and so distributing subtasks equally among CPU cores or among nodes does not strongly affect the performance. Therefore, we can calculate the execution time of those subtasks assigned to a single node from a cluster and predict the full execution time of the whole task on that cluster. In our experiment, we choose four nodes with $2$, $4$, $8$ and $16$ vCPU cores, and all with a Memory-to-CPU Core ratio of $2$. We calculate the execution time of $\frac{{N_a}N_{s} \times \#\textbf{vCPU}}{127,512}$ subtasks for each node and choose the largest one as the predicted execution time of the whole calculation on a cluster with $127,512$ vCPU cores and $2\times 127,512$ gigabytes of memory.

\begin{table}[H]
   \caption{Estimated Execution Time for Simulating Bristlecone-70 Circuits Using $127,512$ CPU Cores}
   \label{tab:estimation}
   \small 
   \centering 
   \begin{tabular}{lcccr} 
   \toprule[\heavyrulewidth]\toprule[\heavyrulewidth]
   \textbf{Circuit} & \textbf{Amplitudes} & \textbf{Subtasks p.a.} & \textbf{Computational Resources} & \textbf{Time (s) p.a.} \\ 
   \midrule
   \multirow{5}{*}{Bristlecone-70$\times (1+28+1)$ } 	& $1$ 				& $256$ 	& ($2k$ vCPU, $4k$ GB)s where $k=1,2,4,8$ 		& $9.71$ \\
     													& $256$ 			& $256$ 	& ($2k$ vCPU, $4k$ GB)s where $k=1,2,4,8$ 		& $0.03$ \\
													& $1, 000$ 		& $256$ 	& ($2k$ vCPU, $4k$ GB)s where $k=1,2,4,8$ 		& $0.024$ \\
												 	& $200, 000$ 		& $256$ 	& ($2k$ vCPU, $4k$ GB)s where $k=1,2,4,8$ 		& $0.02$ \\
													& $200, 000$ 		& $256$ 	& ($88$ vCPU, $160$ GB) $\times 1449$ 			& $0.03$ \\
   \midrule
   \multirow{4}{*}{Bristlecone-70$\times (1+32+1)$} 	& $1$ 				& $1024$ 	& ($2k$ vCPU, $4k$ GB)s where $k=1,2,4,8$ 		& $23$ \\
												 	& $1,000$ 			& $1024$ 	& ($2k$ vCPU, $4k$ GB)s where $k=1,2,4,8$ 		& $0.28$ \\
												 	& $10,000$ 		& $1024$ 	& ($2k$ vCPU, $4k$ GB)s where $k=1,2,4,8$ 		& $0.27$ \\
												 	& $10,000$ 		& $1024$ 	& ($88$ vCPU, $160$ GB) $\times 1449$ 			& $0.36$ \\
   \midrule
   \multirow{4}{*}{Bristlecone-70$\times (1+36+1)$}	& $1$ 				& $65536$ 	& ($2k$ vCPU, $4k$ GB)s where $k=1,2,4,8$ 		& $7.62$ \\
												  	& $10$ 				& $65536$ 	& ($2k$ vCPU, $4k$ GB)s where $k=1,2,4,8$ 		& $3.3$ \\
											      	& $100$ 				& $65536$ 	& ($2k$ vCPU, $4k$ GB)s where $k=1,2,4,8$ 		& $2.69$ \\   
											      	& $100$ 				& $65536$ 	&  ($88$ vCPU, $160$ GB) $\times 1449$ 			& $4.56$ \\   
   \midrule
   \multirow{3}{*}{Bristlecone-70$\times (1+40+1)$}  	& $1$ 				& $4194304$ 	& ($2k$ vCPU, $4k$ GB)s where $k=1,2,4,8$ 	& $412.25$ \\
												  			& $5$ 				& $4194304$ 	& ($2k$ vCPU, $4k$ GB)s where $k=1,2,4,8$ 	& $389.43$ \\
												  			& $1$ 				& $4194304$ 	& ($88$ vCPU, $160$ GB) $\times 1449$ 		& $480.17$ \\
   \bottomrule[\heavyrulewidth] 
   \end{tabular}
\end{table}

From the above table, we observe that the more amplitudes we calculate, the less the execution time per amplitude. This is due to a more balanced computational load for larger $\frac{{N_a}N_{s} \times \#\textbf{vCPU}}{127,512}$. When $N_a$ is large enough, the run time per amplitude will remain relatively stable. We also observe that even when using same number of total vCPU cores, execution time will be slightly reduced when we use a larger number of smaller ECS instances.

\begin{table}[H]
   \caption{Benchmarking Results for Simulating Bristlecone-70 Circuits on Alibaba Cloud}
   \label{tab:benchmarking}
   \small 
   \centering 
   \begin{tabular}{lcccr} 
   \toprule[\heavyrulewidth]\toprule[\heavyrulewidth]
   \textbf{Circuit} & \textbf{Amplitudes} & \textbf{Subtasks p.a.} & \textbf{Computational Resources} & \textbf{Time (s) p.a.} \\ 
   \midrule
   \multirow{1}{*}{Bristlecone-70$\times (1+28+1)$ } & $200,000$ & $256$ & (88 vCPU, 160GB)x1449 & $0.04$ \\
   \midrule
   \multirow{1}{*}{Bristlecone-70$\times (1+32+1)$} & $1,000$ & $1024$ & (88 vCPU, 160GB)x1449 & $0.43$ \\
   \midrule
   \multirow{4}{*}	{Bristlecone-70$\times (1+36+1)$}	& $10$ & $65536$ & (88 vCPU, 160GB)x1449 & $7.6$ \\
		  												 	& $10$ & $65536$ & (88 vCPU, 160GB)x1449 & $6.6$ \\
   															& $100$ & $65536$ & (88 vCPU, 160GB)x1449 & $5.6$ \\
   															& $100$ & $65536$ & (88 vCPU, 160GB)x1449 & $5.9$ \\
   \midrule
   \multirow{1}{*}{Bristlecone-70$\times (1+40+1)$} & $1$ & $4194304$ & (88 vCPU, 160GB)x1449 & $580.7$ \\
   \bottomrule[\heavyrulewidth] 
   \end{tabular}
\end{table}

Let $\ket{\psi}$ denote the output of a random quantum circuit. It is known that the distribution of measurement probabilities $p(x_j)=\vert \bra{x_j}\psi\rangle\vert^2$ approaches the exponential form $Ne^{-Np}$, known as the Porter-Thomas distribution. Based on the $200,000$ amplitudes we calculated for Bristlecone-70 circuits with depth $1+28+1$, we plot the distribution of $Np$, which closely matches the Porter-Thomas form.

\begin{figure}[h!]
\centering
  \caption{The distribution of measurement probabilities, which closely matches the Porter-Thomas form $Ne^{-Np}$.}
  \includegraphics[scale=0.65]{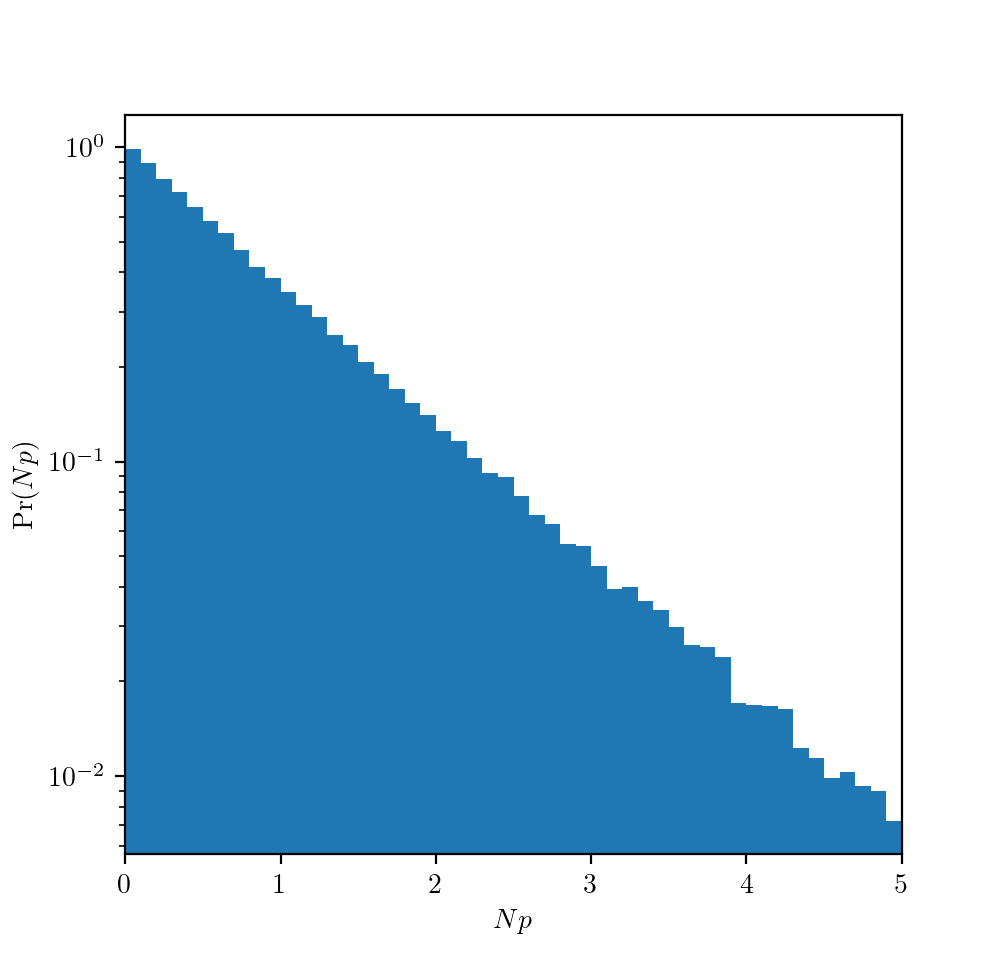}
\end{figure}

\section*{Acknowledgements}
We would like to thank our colleagues from various teams in Alibaba Cloud Intelligence supporting us in the numerical experiments presented in this paper.  We thank Jianxian Zhang and his team's support from the local data center and Jiaji Zhu and his team for informing us on progress in PanGu 2.0, the high performance distributed file system, which will enhance future deployment.

\bibliographystyle{plain}
\bibliography{main}
\end{document}